\documentclass[aps,prd,nofootinbib]{revtex4}
\usepackage[caption=false]{subfig}

\usepackage[T1]{fontenc}
\usepackage[utf8]{inputenc}
\usepackage{lmodern}
\usepackage{amsmath}
\usepackage{amsfonts}
\usepackage{booktabs}
\usepackage{graphicx}
\usepackage{todonotes}
\usepackage[english]{babel}
\usepackage{placeins}

\usepackage{tikz}

\usepackage{lineno}
\usepackage{hyperref}
\graphicspath{ {figures/}}
\modulolinenumbers[5]

\begin{document}

%\begin{frontmatter}

\title{CDT and Cosmology}
\date{\today}
\author{L. Glaser and R. Loll}
\address{Radboud University,
Institute for Mathematics, Astrophysics and Particle Physics,
Heyendaalseweg 135, NL-6525 AJ Nijmegen, The Netherlands.}

\begin{abstract}
\noindent In the approach of Causal Dynamical Triangulations (CDT), quantum gravity is obtained as a scaling limit of a
non-perturbative path integral over space-times whose causal structure plays a crucial role in the construction.
After some general considerations about the relation between quantum gravity and cosmology,
we examine which aspects of CDT are potentially interesting from a cosmological point of view, focussing on the
emergence of a de Sitter universe in CDT quantum gravity.
\end{abstract}

%\begin{keyword}
%\texttt{elsarticle.cls}\sep \LaTeX\sep Elsevier \sep template
%\MSC[2010] 00-01\sep  99-00
%\end{keyword}
%\end{frontmatter}
\maketitle
%\linenumbers
\tableofcontents
\section{Introduction}
\label{intro}
\noindent The aim of cosmology is to describe the behaviour of the universe as a whole. The current standard model of cosmology ($\Lambda$CDM)
assumes that gravity on cosmological scales is described by general relativity (GR) or, more correctly, the Friedmann equations. The latter
can be derived from Einstein's field equations by assuming from the outset that the spatial universe is homogeneous and isotropic.
This is an extremely strong symmetry assumption, describing a spatial universe where
all sources of gravity are smeared out evenly and which is devoid of any local structure.
It is clearly not satisfied by the real universe at the scale of stellar systems, galaxies, galaxy clusters or the even larger galaxy
filaments. However, astrophysical observations appear to be compatible with {\it statistical} homogeneity and isotropy at a scale of about 100 Mpc
and larger.
In this sense, Friedmann-Lemaitre-Robertson-Walker (FLRW) space-times
serve as effective models of gravity on sufficiently large scales. This leaves aside the question of how to average
in an intrinsically nonlinear theory like general relativity, as would be needed to gain a more detailed understanding of the relation between the
full theory and the real, large-scale universe approximated by a spatially flat FLRW geometry.

Should we be surprised by how well and robustly the universe at large can be described by a FLRW space-time, where
at each moment $t$ the entire dynamical content of the metric tensor $g_{\mu\nu}(t,x)$ is captured by
a single number, the global scale factor $a(t)$? This question is not easy to answer, given that -- despite the tremendous successes of
standard cosmology -- some of its
key fundamental issues remain unresolved, including the nature of dark matter and the cosmological constant problem (see \cite{dark} for a recent review).
What could we possibly learn about cosmology from a theory of quantum gravity, that we could not learn
from classical GR alone?

Note first that even if general relativity (and not some
version of ``modified  gravity'') is the correct theory describing gravity on very large scales,
the geometrical symmetry reduction to the FLRW setting does not have fundamental status, since the exact symmetry
is not realized in nature at any scale. This implies that a theory of quantum gravity, even if it has GR as its classical limit,
need not reproduce the $\Lambda$CDM-model as such, but should only be able to explain the observed large-scale features of the
universe, including those that are described correctly by the $\Lambda$CDM-model. Secondly, because of the
non-renormalizability of perturbative quantum gravity, we know that an
effective field theory description must break down at sufficiently large energies, beyond which it should be ``ultraviolet completed''.
This motivates the search for {\it non-perturbative quantum gravity}, a fundamental quantum theory of gravity valid on {\it all} scales,
which includes a specification of its degrees of freedom -- geometric, pre-geometric or otherwise -- and their dynamics at the Planck scale.

All quantum and classical aspects of gravity and cosmology must ultimately follow from this fundamental, ``true'' theory of gravity,
including the quantum behaviour of the very early universe and finding the non-perturbative quantum ground state of the universe.
There may also exist less obvious
implications of the quantum theory for {\it large}-scale cosmology, for example,
in the form of quantum-gravitational contributions to the observed cosmological constant, implying some ``mixing'' of ultraviolet and
infrared excitations. In any event, true cosmology should be derived from a consistent, non-perturbative quantum description of
matter and gravity beyond classical GR. Of course, even a fully-fledged theory of quantum gravity is likely to leave some questions
unanswered, for example, those of the initial conditions of our universe.

Given the robustness of the FLRW paradigm and the $\Lambda$CDM-model at the classical level, one may expect
that true cosmology can be cast in similar terms. However, in view of the conundrums of standard cosmology, there must be
at least {\it some} surprises and new insights awaiting us, which from our current viewpoint are unobvious, perhaps run counter
to folklore or look incompatible with standard assumptions. When we examine candidate theories for quantum gravity -- which
at present are more or less incomplete -- and their possible implications for cosmology, we need to watch out for such features
and to some degree be open-minded.

\begin{figure}
\centering
\includegraphics[width=0.6\textwidth]{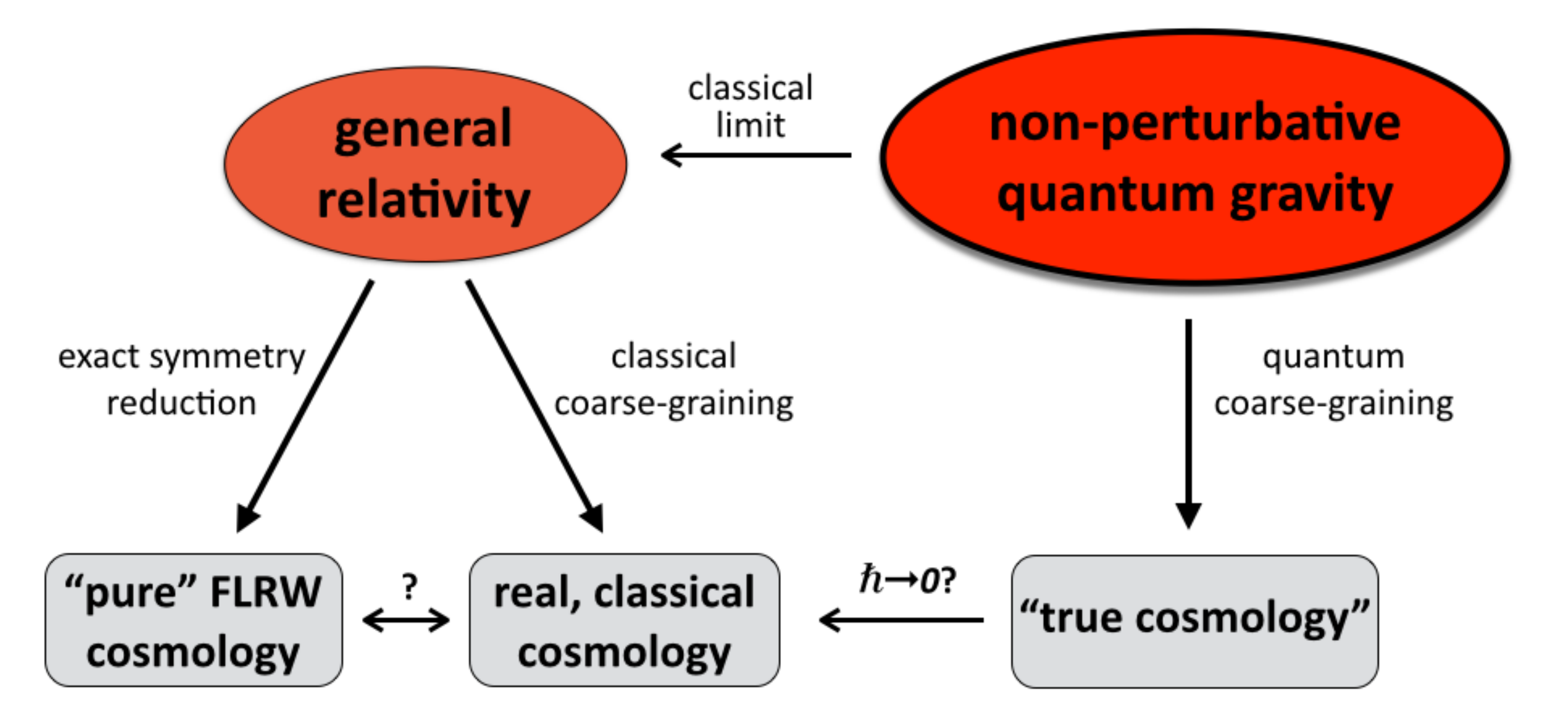}
\caption{\label{fig:GRQGcosmo} Different ways to arrive at cosmology from general relativity and non-perturbative quantum gravity.}
\end{figure}

Fig.\ \ref{fig:GRQGcosmo} illustrates the relation between the theories of quantum gravity, general relativity and cosmology.
Assuming that GR is
the correct classical theory of gravity on length scales much larger than the Planck scale $\ell_{\rm Pl}=1.6\times 10^{-35}m$,
its exact reduction to the FLRW-sector occurs by imposing homogeneity and isotropy on
spatial slices of constant time $t$. By contrast, what we observe astrophysically is a universe where this symmetry
is realized (at best) in a statistical sense and on length scales above 100 Mpc. In standard cosmology, we nevertheless interpret
these data in terms of a FLRW model.
For which observables and on what scales this procedure is justified and yields a good quantitative description
is the subject of ongoing analytical and numerical research.
A main difficulty is how to evaluate in GR the so-called ``backreaction'' effect of inhomogeneities on smaller scales
on the dynamics of the universe on larger scales, in a situation
where neither averaging nor coarse-graining are uniquely defined processes (for recent overviews,
see e.g. \cite{challenges,inhomogeneous} and references therein).

Interestingly, similar issues arises
when considering candidate theories of non-perturbative quantum gravity that provide concrete models of microscopic dynamics at the
Planck scale, typically in a regularized form involving some kind of discretized ``building blocks''.
To demonstrate that such a theory correctly reproduces general relativity in a classical limit or to extract cosmological
predictions one needs to study quantum observables describing physical properties on large scales. To achieve this, many orders of
magnitude need to be bridged, by ``coarse-graining'' observables and/or ``integrating out'' microscopic degrees of freedom.
This calls for a non-perturbative and background-independent version of the renormalization group, to determine how physics
depends on the scale, analogous to what is done in the approach of ``Asymptotic Safety'' based on the continuum metric $g_{\mu\nu}(x)$
(see \cite{as} for an assessment of its cosmological implications).
However, the physical principles underlying such a construction and its detailed numerical implementation have
not yet been investigated widely, in part hampered by the current incompleteness of quantum gravity theories and
the scarcity of effective numerical tools.

In the remainder of this article we will focus on the approach of Causal Dynamical Triangulation (CDT), and examine more closely
what we have learned in practice about some of the theoretical issues just raised. To what extent
can one rederive aspects of standard cosmology from the full, non-perturbative quantum theory?
What are the obstacles and how robust are the results?
Has one uncovered new quantum phenomena at the Planck scale, which may eventually lead to predictions that can be tested
in cosmology? What more needs to be done to complete our theoretical understanding and bridge the gap to phenomenology?

In Sec.\ \ref{cdtintro} we recap some important ingredients and properties of CDT quantum gravity, and in Sec.\ \ref{couplings}
we sketch what is known about the model's phase structure. The main part of our discussion on CDT and cosmology is
contained in Sec.\ \ref{global}. After describing possible pathologies of the non-perturbative path integral, we focus on
the emergence of a de Sitter universe, which is a key result in CDT. We explain how it comes about dynamically, what
we know about its physical properties, and that there is still some way to go to derive hard, quantitative predictions about true
cosmology from Causal Dynamical Triangulations. Our treatment is mostly non-technical, with plenty of references
for those wanting to delve deeper into the subject.

\section{CDT Quantum Gravity}
\label{cdtintro}

\noindent Quantum Gravity in terms of Causal Dynamical Triangulation provides an explicit realization of a non-perturbative,
Planckian quantum dynamics, obtained as the continuum limit of a regularized path integral.
Importantly, it comes with powerful numerical tools that have made it possible to extract
certain large-scale properties of the ``ground state of the universe'' and to compare them to a classical universe of
FLRW-type. It is not our intention here to go into the details of the technical construction of the CDT path integral, which
have been given in many places (see e.g. the review \cite{physrep}), but only to highlight
some properties of the construction and to describe briefly some of its results that have some bearing on the theme of
``cosmology from quantum gravity''. In this spirit, we will cover the theory in four space-time dimensions only, and only
the case of pure gravity, because it is the one most investigated and far from trivial. Coupling matter is technically
straightforward and has been studied extensively in Dynamical Triangulations (DT), the precursor and counterpart with
Euclidean signature of CDT \cite{4dsimpl}. However, identifying physically interesting matter-gravity observables is not easy and will
not be discussed further here, although it is an important area of current and future research. A point particle coupled to
CDT was considered in references \cite{point} and \cite{wilson}.

In a nutshell, CDT quantum gravity provides a well-defined regularized version of the formal
path integral over space-time geometries $g$, the famous ``sum over histories''
\begin{equation}
Z=\int {\cal D}g\, {\rm e}^{i S^{\rm EH}[g]},
\label{formalPI}
\end{equation}
where each space-time $g$ is associated with a complex amplitude depending on the Einstein-Hilbert action evaluated at $g$.
From the regularized path integral, one then seeks to obtain a quantum gravity theory in a suitable scaling limit, as the regulators are removed.
An important property of this construction, if such a limit can be shown to exist, is that the final continuum theory will be
largely independent of the details of the regularization, a phenomenon known as {\it universality}. It also means that there are few
adjustable parameters, which is an enormous bonus if one is interested in robust results and the potential predictive power of
the theory.
The entire construction takes place within the realm of quantum field theory, without invoking any exotic ingredients,
and can be thought of as the gravitational analogue of lattice gauge theory.
Also the motivation is similar to that of QCD, say, in the sense that through numerical (Monte Carlo) methods
one hopes to gain nontrivial and quantitative information about the non-perturbative regime of the theory, which currently is not
obtainable by other means. Of course, quantum chromodynamics is a theory we already know much more about than quantum gravity.

Another difference with QCD are the symmetries of the theory. The freedom to perform
space-time diffeomorphisms (a.k.a. coordinate transformations) in general relativity is an inevitable feature of
the standard continuum formulation. Already classically, this redundancy obscures the distinction between what is physical and
what is a mere coordinate effect. The situation in the quantum theory is usually worse,
since beyond the linearized theory it is very difficult to account for the full diffeomorphism symmetry in a consistent manner,
especially in the presence of discreteness (fundamental or cutoff), and to control the associated ambiguities and infinities.
A beautiful and perhaps underappreciated feature of quantum gravity in terms of dynamical triangulations
is the fact that this problem is {\it absent}. The reason is that the configuration space of the regularized path integral is
a space of triangulations, given by piecewise flat manifolds made of equilateral triangular building blocks of geodesics edge length $a$.
It is based on the ingenious idea of Regge's \cite{regge} to describe the curved space-times of GR in terms of purely geometric data,
namely, the edge length assignments and connectivity (``gluing'') data of triangulated manifolds,
thereby completely avoiding the introduction of coordinates and their associated redundancy.

\begin{figure}
\centering
\includegraphics[width=0.5\textwidth]{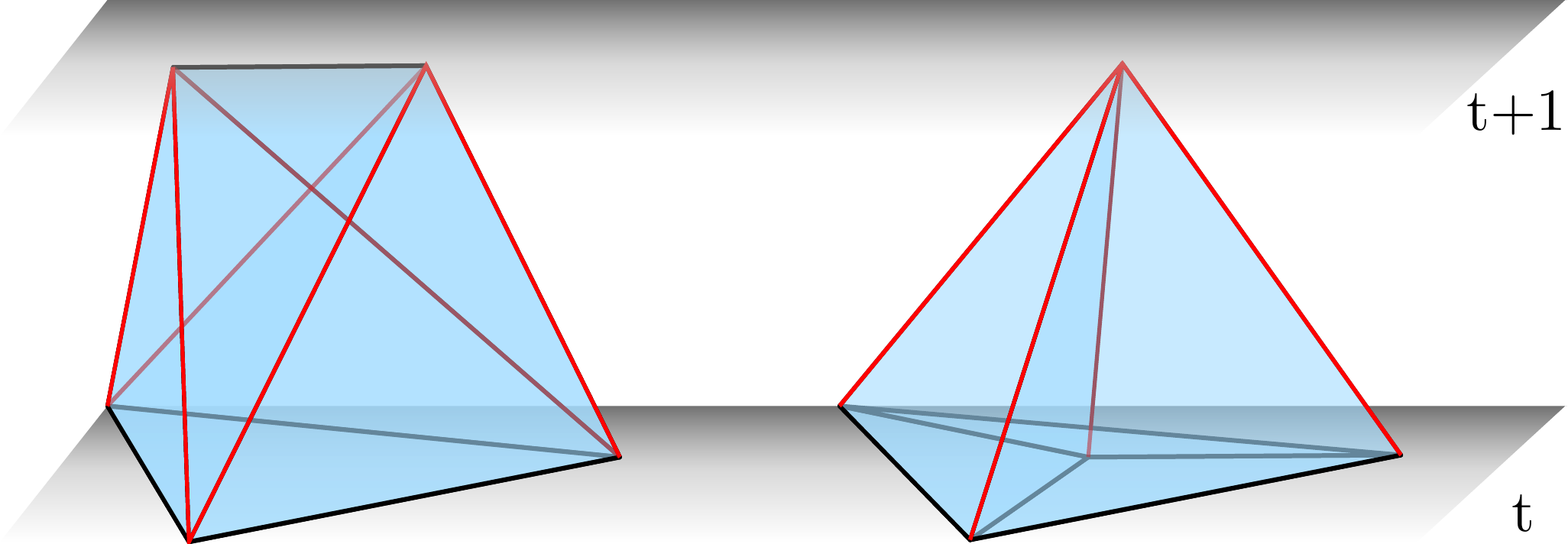}
\caption{\label{fig:simplices}The two types of Minkowskian four-simplex that serve as the elementary building blocks of 4D CDT, the
(3,2)-simplex (left) and the (4,1)-simplex (right). They differ
in their assignments of space- and time-like edges, see \cite{physrep} for details. }
\end{figure}
Fig.\ \ref{fig:simplices} shows the
two kinds of four-simplices used to construct the space-time configurations summed over in the CDT path integral.
The reason why one type of building block is not sufficient has to do with the fact that in Lorentzian signature a four-simplex
cannot be equilateral, since in that case the flat geometry induced on its interior would be Euclidean and not Minkowskian.
However, we can make the four-simplices ``as equilateral as possible'' by using two different edge lengths, in the case of CDT chosen to be
space-like and time-like. In terms of the UV-cutoff $a$, the length-squared in CDT of
all space-like edges is $\ell_s^2=a^2$, and that of all time-like edges $\ell_t^2=-\alpha a^2$, for some fixed $\alpha >0$.
The figure also illustrates another important aspect of the standard CDT set-up, namely, that the elementary building blocks are
by construction arranged in layers that can be labelled by a discrete ``time'' parameter $t$. The geometry of the spatial universe
at any integer value $t$, shared by two consecutive space-time layers of ``height'' $\Delta t\! =\! 1$,
is a piecewise flat Riemannian space consisting of equilateral tetrahedra with edge length $\ell_s$.
Note that $t$ is merely one of the parameters characterizing the regularized geometries, and not related to any
gauge choice, since there are no coordinates in the first place\footnote{nor would it
in general be possible to introduce coordinates in neighbourhoods that extend beyond individual building blocks, because of the presence of
curvature singularities}.
Whether or not $t$ acquires any physical meaning in a suitable continuum limit cannot be determined a priori.
Anticipating our discussion of Sec.\ \ref{global} below, in the CDT phase where a de Sitter-like behaviour is found, a
renormalized version of $t$ attains the status of cosmological proper time. Whether there are observables
with respect to which (a continuum version of) $t$ can be interpreted in terms of a more local notion of time is currently unknown.

At the regularized level, the presence of a ``time'' $t$ allows us to implement in a simple way the causality restriction that
puts CDT quantum gravity in a different universality class from its purely Euclidean DT counterpart, which is the requirement
that the topology of {\it space} of any path integral configuration should not change in time. Of course, such topology changes
are never permitted in the classical
theory because they are associated with degeneracies of the Lorentzian metric $g_{\mu\nu}$.
A major insight gained from CDT quantum gravity is that such restrictions on the causal structure
are also essential in the non-perturbative quantum theory to obtain a theory with a good classical limit.
The standard choice made for the topology of the spatial slices is a that of a three-sphere $S^3$. A first study
using tori $T^3$ instead of spheres as spatial slices has appeared recently \cite{torus}.

Imposing causality conditions can in principle be done without referring to an explicit time slicing, and leads to a slightly
generalized version of CDT in terms of ``Locally Causal Dynamical Triangulations (LCDT)''. This variant has been implemented numerically
in three space-time dimensions, where it appears to lead to large-scale results equivalent to those of CDT \cite{LCDT1,LCDT2},
providing nontrivial evidence that CDT and LCDT lie in the same universality class.
However, apart from being more complicated and numerically expensive, it is unclear whether this model is still unitary.
Unlike CDT quantum gravity \cite{cdt2001,physrep}, LCDT does not have a transfer matrix and does not
obviously obey reflection positivity, an important but sometimes neglected property of
lattice gravity models and a prerequisite for recovering unitarity in the continuum limit.

Let us finally note another exceptional feature of CDT quantum gravity, which is the existence of a well-defined ``Wick rotation'', which takes
the form of an analytical continuation of one of the parameters describing the regularized, piecewise flat geometries (more specifically,
a continuation $\alpha\mapsto -\alpha$ through the lower-half complex plane of the parameter $\alpha$ introduced above).
It maps each
curved Lorentzian CDT space-time to a unique curved Euclidean space-time, while simultaneously transforming the corresponding complex
weight factor exp($iS$) to a real Boltzmann factor exp($-S_{\rm E}$) depending on the Euclideanized action $S_{\rm E}$,
which is essential for being able to apply Monte Carlo simulations.
Note that there is no known continuum analogue of this Wick rotation for general metrics or diffeomorphism equivalence classes of metrics.

The unique combination of properties of the CDT path integral just sketched has made it possible to derive a number of new and
unexpected results in non-perturbative quantum gravity. In what follows, we will focus on those that are potentially relevant from the
point of view of cosmology and describe them briefly in turn.

\section{Couplings and phases of CDT}
\label{couplings}

\noindent The first step in understanding the dynamical system defined by the CDT-implementation of the path integral (\ref{formalPI})
is an investigation of its behaviour as a function of the values of the bare coupling constants of the model, which
are the parameters in the gravitational action that can be varied freely. After Wick rotation, the Euclidean version of the bare Einstein-Hilbert action
evaluated on a piecewise flat manifold $T$ -- again following Regge \cite{regge} -- reads \cite{reconstructing,physrep}
\begin{align}
S_{E}(T)=-(\kappa_0 +6 \Delta ) N_0(T) +\kappa_4 (N_4^{(4,1)}(T) +N_4^{(3,2)}(T))+\Delta (2 N_4^{(4,1)}(T)+N_4^{(3,2)}(T)),
\label{sbare}
\end{align}
where $N_0(T)$ counts the number of vertices in the triangulation $T$, and $N_4^{(3,2)}(T)$ and $N_4^{(4,1)}(T)$ the numbers of the
two types of four-simplices introduced previously in Sec.\ \ref{cdtintro}. The coupling $\kappa_0$ is proportional to the
inverse bare Newton constant, and $\kappa_4$ is essentially equal to the bare cosmological constant. In line with the standard philosophy of
critical phenomena that underlie the CDT construction, the bare, dimensionless couplings do not have a direct physical meaning,
but only their renormalized, dimensionful
counterparts will. Whether a meaningful continuum theory depending on such renormalized couplings exists depends on whether the
underlying statistical model possesses interesting
scaling limits in the vicinity of critical points in the phase diagram, generally found along lines of phase transition of second or higher order.

The phase diagram of CDT quantum gravity is spanned by three bare parameters. In addition to $\kappa_0$ and $\kappa_4$, a third, so-called
``asymmetry parameter'' $\Delta$ becomes relevant when discussing possible continuum limits.
It is called asymmetry parameter because it is a function of $\alpha$, which we introduced above as the ratio between
the length of space-like and time-like edges in the triangulated geometries. The reason for using $\Delta$ instead is that the bare action
(\ref{sbare}) depends on it in a simple way. Note that $\Delta$ vanishes when all edges of the four-simplices
(in the Wick-rotated geometries) have equal length,
i.e. $\Delta (\alpha\! =\!  1)=0$. At the level of the bare action, different values of $\Delta$ amount to trivial, finite field rescalings.
However, in the non-perturbative regime of CDT where interesting physics is observed, its character is changed to that
of a relevant parameter. Its precise physical meaning must be extracted by measuring suitable observables near
phase transitions.
Whether or not (a renormalized counterpart of) the coupling $\Delta$ is related to time-space anisotropy of some continuum theory,
perhaps in the spirit of Ho\v rava-Lifshitz gravity \cite{horava,cdthl}, remains to be understood. There are currently too few
non-perturbative observables to make a specific statement about any genuinely physical anisotropy, for example, regarding
the large-scale behaviour of the quantum geometry.

Before discussing details of the phase diagram, note that {\it CDT predicts a positive (renormalized) cosmological constant $\Lambda$}!
This generic feature of DT models has to do with the convergence behaviour of the non-perturbative path integral and the
additive renormalization the cosmological constant undergoes in the limit as the number of building blocks is taken to infinity.
CDT does not predict any particular value for the physical cosmological constant (or any other coupling constant), only that
it should be positive, which at present is of course compatible with our understanding of the universe, based on $\Lambda$CDM cosmology.

\begin{figure}
\centering
\includegraphics[width=0.5\textwidth]{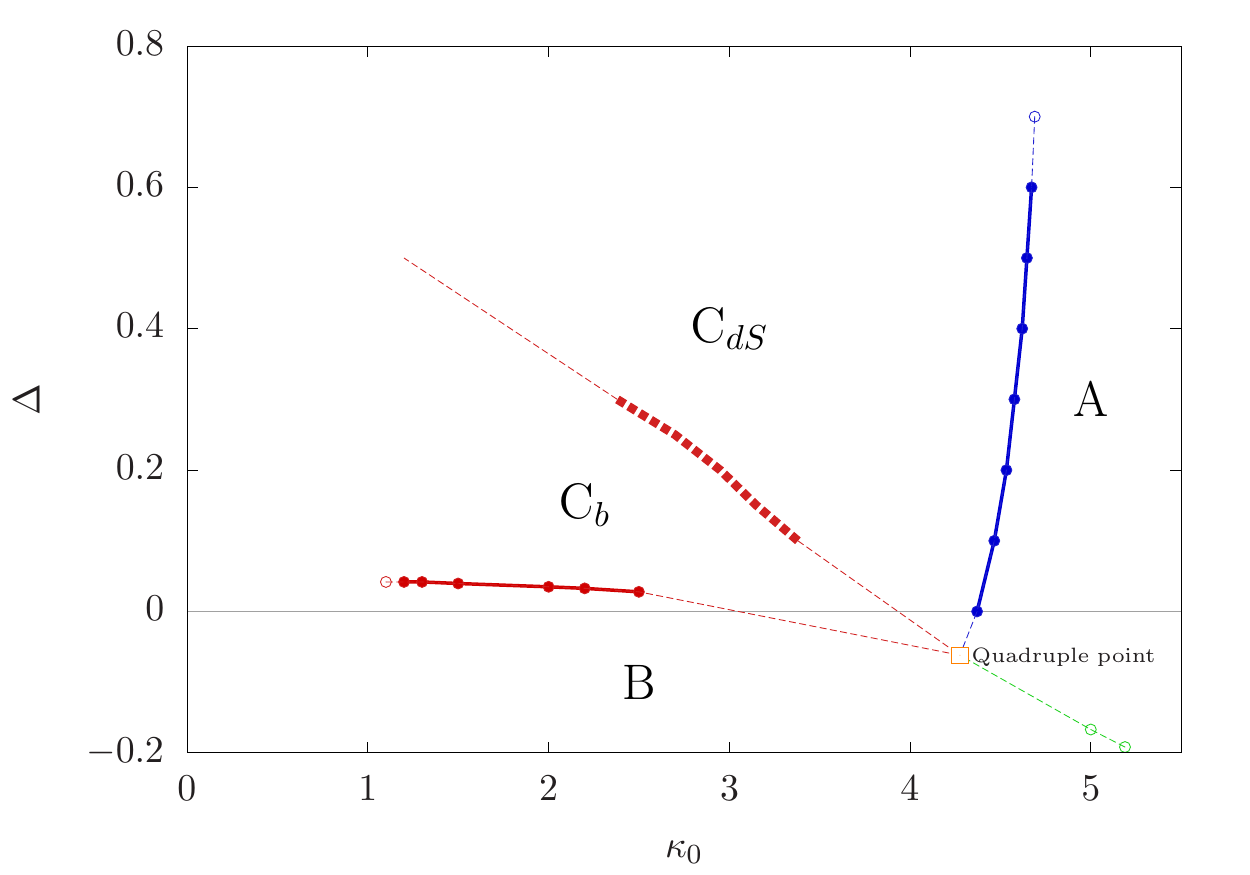}
\caption{\label{fig:phases} Phase diagram of CDT quantum gravity, with phase transition lines (from \cite{characteristics}).
The fat dots and squares refer to actual measurements, the rest is based on extrapolation, including the hypothesized quadruple point.}
\end{figure}
The phase diagram is investigated with the help of Monte Carlo simulations (see~\cite{emergence,physrep} for more information),
which for technical reasons are performed at fixed four-volume, implemented by adding a volume-fixing term to the action
and tuning $\kappa_4$ to its pseudo-critical value. The dependence on the remaining two bare parameters $\kappa_0$ and
$\Delta$ is shown in the phase diagram of Fig.\ \ref{fig:phases}.
Remarkably, and so far unique in non-perturbative quantum gravity theories, {\it CDT has at least one second-order transition line},
the one separating phases $B$ and $C_b$ \cite{transition1,transition2}. As mentioned above, the existence of such transition points
is a crucial prerequisite to defining scaling limits of the theory. A new transition line,
subdividing the old $C$-phase into the de Sitter phase $C_{dS}$ and the bifurcation phase $C_b$,
was discovered only recently \cite{effective,signature}
and is also a candidate for a higher-order transition \cite{newphasetransition}. By contrast, the $A$-$C_{dS}$ line has been
shown to be of first order \cite{transition2}.

One physical criterion to distinguish between the various phases
is the behaviour of the expectation value $\langle V_3(t) \rangle$ of the spatial volume $V_3$ of the quantum universe as a
function of time $t$ \cite{emergence,reconstructing,semilimit}. Volume profiles compatible with a good classical limit are only
found in phases $C_b$ and $C_{dS}$, while the phases $A$ and $B$ seem to be unphysical and are likely to have no continuum
interpretation in terms of general relativity.
This is also corroborated by studying other observables, like the dynamical dimensions
considered below in Sec.\ \ref{global}, which show that an extended, four-dimensional universe emerges on large scales only in the $C$-phases.
The distinction between the bifurcation phase $C_b$ and the de Sitter phase $C_{dS}$ was first spotted in the transfer matrix approach
\cite{transfer}, where one simulates only a pair of adjacent time slices instead of configurations that extend over 80 time steps, say,
and contain the universe in its entirety.
The bifurcation phase is characterized by the appearance of vertices of high order, where many four-simplices share the same vertex.
This leads to a substructure of the quantum geometry, whose physical meaning is currently being investigated and has been tentatively
associated with a breaking of spatial homogeneity, which would be very interesting from a cosmological point of view \cite{characteristics}.
Which of the de Sitter-like properties of the geometry in phase $C_{dS}$ -- described in more detail in the next section -- continue to hold
throughout the phase $C_b$ is an interesting question and the subject of ongoing research.

\section{Recovering cosmology}
\label{global}

\noindent Going back to our considerations of the Introduction, what we have learned from studying non-perturbative models of quantum
gravity, and in very explicit terms from DT and CDT, is that a generic ``sum over histories'' does not produce anything
like a FLRW universe on any scale. In other words, the robustness of a FLRW-like description
of large-scale geometry observed in classical cosmology does not carry over immediately to the quantum context,
where Planckian curvature fluctuations are large. It turns out that the latter are associated with generic, ``entropic'' instabilities,
where the path integral is dominated by highly degenerate configurations that never ``average out'' in expectation values to
produce recognizable semi-classical behaviour. This undesirable outcome is largely insensitive to the choice of microscopic building
blocks, the detailed form of the measure and the gravitational action used (see \cite{measure} for a recent example)
and the presence or otherwise of matter. A well-known type of
pathological behaviour found in the continuum limit of quantum-gravitational systems is associated with so-called
branched polymers, which can be identified in terms of their critical exponents.
They are found in 4D Euclidean DT quantum gravity
for sufficiently large $\kappa_0$ \cite{4dEDT},
but are also present generically in DT in dimension larger than two \cite{3dEDT,5dEDT,10dEDT},
as well as in tensor models \cite{melon}, illustrating the strength of the underlying effect.

The nature of these pathological phases is pre-geometric, in the sense that they are dominated by
specific dynamical geometric modes, to such a degree that a four-dimensional extended universe is never formed.
For example, the branched-polymer behaviour just discussed is usually attributed to a dominance of the local conformal mode,
and leads to a ``quantum geometry'' with a Hausdorff dimension of 2 and a spectral dimension of 4/3.
The challenge is therefore to find quantum gravity models which at least in some region of their phase space exhibit behaviour
that on large scales is compatible with general relativity, possibly in combination with a fine-tuning of couplings near a
phase transition.

CDT quantum gravity was invented to tame the observed pathologies of
non-perturbative Euclidean DT models sufficiently to make a physical interpretation possible, without
constraining local curvature degrees of freedom or otherwise tampering with the gravitational dynamics.
Its key finding is that requiring path integral histories to have a well-behaved causal structure (without acausal points)
leads to completely different quantum gravity theories\footnote{in all space-time
dimensions studied so far: two, three and four} than their purely Euclidean counterparts, which have no notion
of either time or causality. This ``causality condition'' in CDT is physically motivated -- because the real world {\it is} Lorentzian -- but by no
means a necessary physical requirement, since individual path integral configurations are not in themselves physical.
The CDT ansatz has proved very successful since its inception in \cite{firstcdt}, and has given us the first explicit (and so far unique)
example of the {\it emergence of a specific classical, cosmological space-time from a fully non-perturbative and
background-independent quantum gravity theory}, as we will now explain in more detail.

\begin{figure}
\centering
\includegraphics[width=0.6\textwidth]{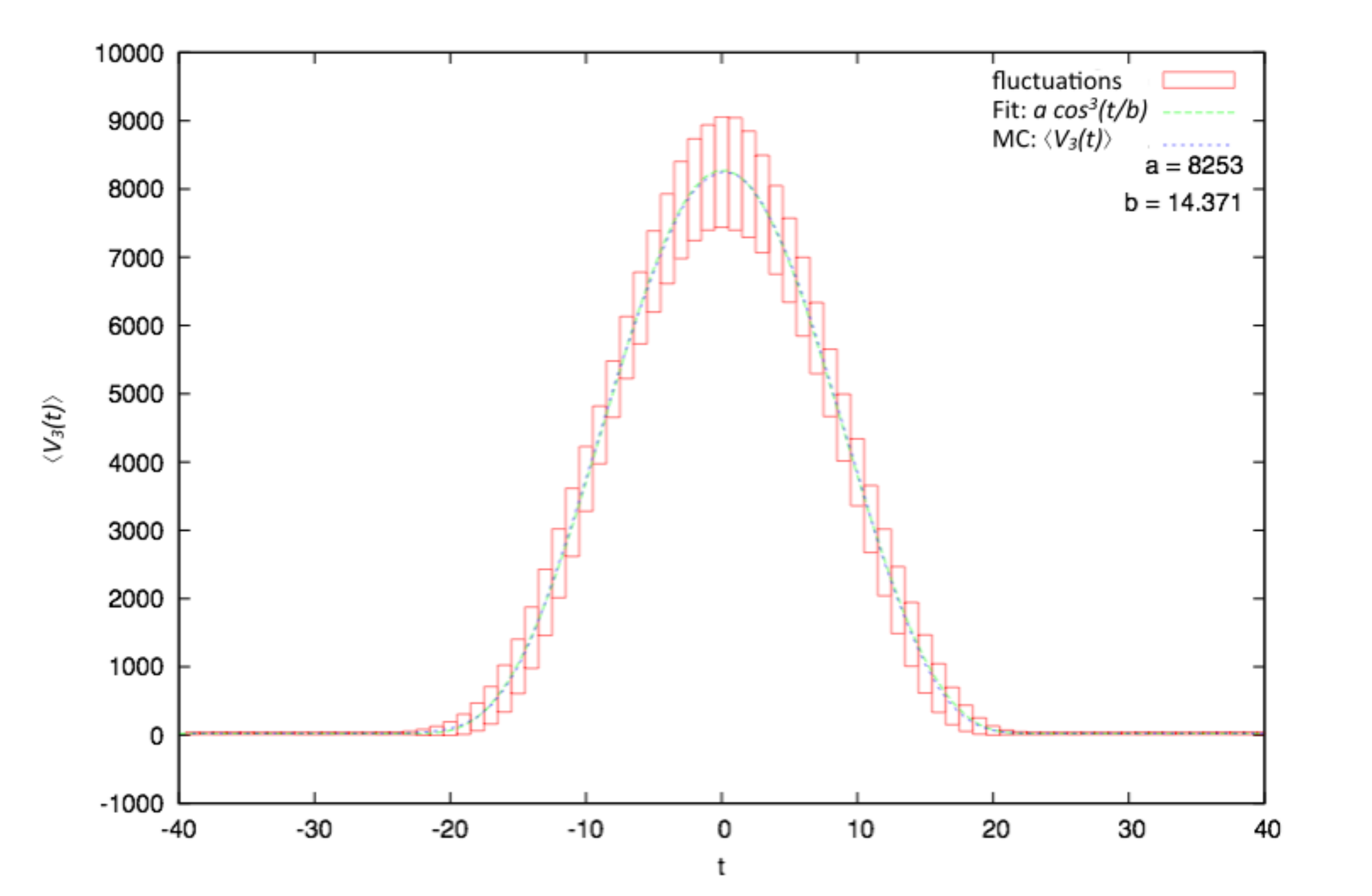}
\caption{\label{fig:desitter}
Monte Carlo measurements of the volume profile $\langle V_3(t)\rangle$
as a function of discrete time $t$, measured in the de Sitter phase $C_{dS}$
at $(\kappa_0,\Delta)=(2.2,0.6)$ and four-volume $N_4=362k$.
A best fit with a cos$^3$-profile derived from the classical cosmological solution with metric
(\ref{dSline}) yields a curve which at the given resolution is essentially indistinguishable from the Monte Carlo data.
The red bars indicate the average size of quantum fluctuations at given $t$. They are large, compared to
the volumes themselves, since the universe is very small, as explained in the text.}
\end{figure}

The observables that enable us to match properties of the non-perturbative vacuum state of pure quantum gravity with
those of a four-dimensional de Sitter space are the volume profile $V_3(t)$ introduced earlier, and two different notions
of dimension. Let us start by describing some results on the former.
To be able to compare with Monte Carlo measurements of the Wick-rotated path integral, consider Euclidean general relativity in
the presence of a positive cosmological constant, and {\it assume} spatial homogeneity and isotropy, as is usually done
in deriving cosmological space-times of FLRW type. This reduces
the Einstein equations to ordinary differential equations of the scale factor $a(\tau_E)$, whose solution is the (Euclidean) de Sitter universe.
Its line element is given by
\begin{equation}
ds^2=d\tau_E^2+a(\tau_E)^2\, d\Omega^2_{(3)}=d\tau_E^2+c^2\cos^2 \left(\frac{\tau_E}{c}\right) d\Omega^2_{(3)},
\label{dSline}
\end{equation}
where $d\Omega^2_{(3)}$ denotes the line element of the unit three-sphere, $c$ is inverse proportional to the square root of the
cosmological constant
and $\tau_E=i\tau$ is Euclidean proper time. From the metric (\ref{dSline}) it is straightforward to compute the associated classical
volume profile as
$V_3(\tau_E)\propto \cos^3 (\tau_E/c )$.\footnote{Note that after a Wick rotation $\tau\mapsto\tau_E$, Lorentzian de Sitter
space becomes Euclidean de Sitter space, which geometrically is a round four-sphere.} As illustrated by Fig.\ \ref{fig:desitter},
the measured average volume profile of the dynamically
generated quantum universe in CDT matches this functional form perfectly \cite{desitter1,desitter2,semilimit}. Moreover,
the data from the computer simulations can be used to reconstruct the effective action for the three-volume
that leads to the measured volume profile $\langle V_3(t)\rangle$.
This can be done from simulations of the full universe \cite{desitter2} and from the transfer matrix system that uses only two time
slices \cite{effective},
and in both cases matches the corresponding classical minisuperspace action up to an overall sign (more about which below).
Even more, the behaviour of the quantum fluctuations $\delta V_3(t)$ of the three-volume has also been measured and compared to a
perturbative continuum calculation around de Sitter space (in quadratic approximation), yielding a good agreement of the
low-lying eigenmodes \cite{desitter2}.

This ``recovery of cosmology''  is a remarkable result. Without putting in any preferred background geometry from the start, there is
a whole region in the phase space of CDT quantum gravity where the ``democratic'' superposition of all histories produces
a quantum geometry whose shape matches that of a classical FLRW universe. The same is true for the large-scale {\it dimension}
of this universe, which turns out to be 4 (within measuring accuracy), as required for a classical space-time \cite{emergence,spectral,reconstructing}.
The study of dynamical dimensions in quantum gravity has in the meantime become a whole subject in itself.
The qualifier ``dynamical'' refers to the fact that in a
non-perturbative, Planckian regime -- inasmuch as prescriptions for measuring a dimension continue to be meaningful -- these dimensions
can differ from their canonical value of 4, and need not be integers either.\footnote{One reason why the dimension does not have
to be 4, despite the use of four-dimensional building blocks, is that we perform a nontrivial continuum limit in which specific
properties present at the cutoff scale need not survive.}
A case in point is that of the spectral dimension, which is the dimension
``seen'' by a diffusion process. The fact that quantum space-time seems to undergo a {\it dynamical dimensional reduction} to a value
of or near 2
when one approaches the Planck scale is a phenomenon first discovered in CDT quantum gravity \cite{spectral,reconstructing},
which has since sparked a whole industry of determining spectral dimensions across a variety of quantum gravity approaches
(see e.g. \cite{carlip} and references therein). It is intriguing that many can reproduce evidence of such a dimensional reduction,
but there are several reasons for not jumping to conclusions about the implications of this result.
Firstly, due to the incompleteness of the underlying quantum gravity theories, many derivations require additional ad-hoc assumptions,
which may make the result look more universal than it actually is. Secondly, although the requirement of recovering four-dimensionality
as part of taking a classical limit is nontrivial and useful in identifying unphysical behaviour (like in the case of branched polymers),
the spectral and Hausdorff dimensions that are usually considered provide only a very loose (and again pre-geometric) characterization of
quantum geometry. Non-classical spaces that share the same dimensions can be totally different with regard to other geometric
properties. Thirdly, and most importantly, as long as we do not have a better understanding of coupled gravity-matter systems
on the scales associated with the dimensional reduction, its physical interpretation and potential cosmological implications remain
unclear (see \cite{thermal} for a recent critical assessment from a phenomenological point of view). This is an important and
interesting field for future study.

\begin{figure}[t]
\centerline{\scalebox{0.55}{\rotatebox{90}{\includegraphics{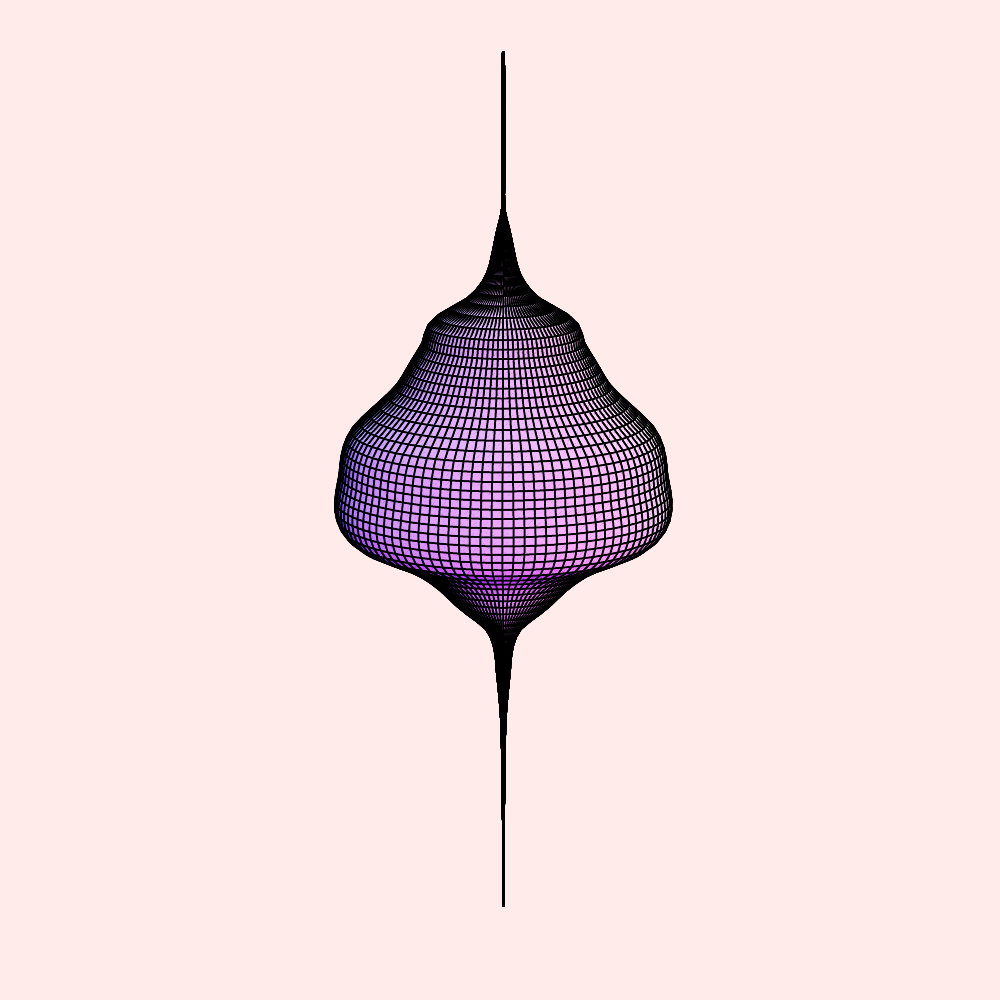}}}}
\caption[phased]{{\small
Monte Carlo snapshot of the volume profile of a typical universe in the de Sitter phase of CDT, at
$(\kappa_0,\Delta)=(2.2,0.6)$, four-volume $N_4\! =\! 91.1k$
and total time extension $t_{total}\! =\! 40$. The time direction lies along the horizontal axis.}}
\label{fig:dspic}
\end{figure}
Returning to the non-perturbative emergence of a de Sitter universe in CDT, some further comments are in order on
the nature of this result.
To begin with, in the relevant region of phase space and for the range of discrete four-volumes $N_4$
explored in the simulations, the {\it physical} size of the de Sitter universe is of the order of 20$\ell_{\rm Pl}$ across.
The identification of lattice with physical units that underlies this number is not ad hoc, but derived from a comparison of the measured spatial
volume-volume correlator with its expected continuum counterpart \cite{desitter1,desitter2}. There is no obstacle in principle
to making the universe larger in terms of physical units, but there are obvious computational limitations to increasing $N_4$
significantly. At any rate, it may be more interesting to go the other way and increase the number of lattice units
$a$ per physical unit\footnote{The lattice
spacing $a$, introduced earlier as the length of an edge of a simplex, should not be confused with the Friedmann scale factor $a(t)$ used elsewhere
in the text. Note that all CDT results quoted are obtained through finite-size scaling in the limit
as $N_4\rightarrow\infty$ or, equivalently,
as the UV-cutoff $a\rightarrow 0$.}, to better resolve quantum effects at the Planck scale. Because CDT dynamics at the cutoff scale -- like in any lattice formulation --
is largely an artefact of the regularization chosen, data points near the cutoff are always excluded when determining observables.
The best resolution achieved in the simulations so far is on the order of $a\sim\ell_{\rm Pl}$, i.e. one lattice spacing per
Planck unit \cite{desitter2,searching}.
This is not quite enough to distinguish between quantum corrections to the minisuperspace action produced by CDT, expected to
become relevant for small values of the scale factor, and discretization ambiguities at the cutoff scale.
In terms of possible applications, quantum modifications at small $a(t)$ were invoked in arguing for a specific form of the
``wave function of the universe'', derivable directly from CDT data \cite{semiclassical}.
Further improvements in the efficiency of the simulations near the phase transition lines are expected to bring measurements
in this short-distance regime within reach.

Next, the snapshot taken from a Monte Carlo simulation of the de Sitter universe shown in Fig.\ \ref{fig:dspic} could be misunderstood as
depicting a universe whose local structure is approximately smooth and close to that of a round four-sphere.
However, {\it nothing could be further from the truth}! What is depicted in the figure is a single global degree of freedom, the
total three-volume as a function of time\footnote{Fig.\ \ref{fig:dspic} is simply a curve made into a body of revolution.}, just like in the graph of Fig.\ \ref{fig:desitter}. This volume variable seems to be very robust,
in the sense that it displays semi-classical behaviour at merely one order of magnitude above the Planck scale.
By contrast, the local geometry of the CDT de Sitter universe is highly singular and quantum fluctuating, and cannot
be captured by any classical metric, not even in an averaged sense, let alone a constantly curved one. Instead, it exhibits a
number of fractal-like properties \cite{reconstructing,geometry}. This means that the dynamically generated quantum universe of
the CDT formulation lives in a truly non-perturbative regime and cannot serve as a classical de Sitter ``background''.
Quantum field theoretic computations on a de Sitter background, which frequently occur in a cosmological context,
cannot be imported in a straightforward way to this non-perturbative setting, unless the corresponding observables
can be reformulated in a diffeomorphism-invariant and background-independent manner.
Of course, this is one facet of the more general challenge in non-perturbative quantum gravity to
identify interesting quantum observables and extrapolate them to a semi-classical regime. One distinct advantage of CDT quantum
gravity in this endeavour is that these questions can be addressed in a well-defined and concrete calculational framework.

By focussing on the dynamics of the three-volume $V_3(t)$ (equivalently, the scale factor $a(t)$) of the quantum universe, we are following the path
of extracting ``real cosmology'' from the full, non-perturbative quantum theory, as depicted in Fig.\ \ref{fig:GRQGcosmo}. In obtaining the
average volume profiles, one is de facto integrating over all other geometric modes, without making any symmetry assumptions
a priori. This should be contrasted with standard quantum cosmology, which amounts to a quantum-mechanical treatment
of only those degrees of freedom that are left after a classical symmetry reduction. (In the gravitational sector of FLRW cosmologies, this is
just $a(t)$.)
Especially in the Planckian regime, where all modes of the metric give potentially large contributions, there is no particular reason why the
two methods should agree. Phrased differently, one would not in general expect ``reduction'' (from the full gravitational field to a
global scale or volume variable) to commute with ``quantization''.

This latter assertion is supported by examining the mechanism behind the emergence of the de Sitter universe in CDT quantum gravity,
which is genuinely non-perturbative and has no analogue in quantum cosmology. The quantum space-time generated by the CDT
path integral is the state which minimizes the ``effective'' Euclidean action that determines the dynamics in that region of phase space.
In the region where a physically interesting de Sitter behaviour is found, this effective action receives significant contributions from
both the bare action and the entropy, by which we mean the number of configurations contributing at a given value of the bare action.
In a statistical mechanics language, this behaviour and the associated phase transitions appear to be ``entropy-driven'', similar to
what happens in the Kosterlitz-Thouless transition of the two-dimensional XY-model (see \cite{entropy,physrep} for related discussions).
We do not have direct analytical access to the full effective action of CDT to illustrate this point explicitly.
However, as alluded to above, an explicit, integrated-out version of this action
for the global conformal mode (a.k.a. the scale factor) has been derived in CDT \cite{desitter2,effective}.
As already mentioned, it can be matched to the standard Euclidean
minisuperspace action of a homogeneous and isotropic universe \cite{haha},
\begin{equation}
S^{mini}_E= \frac{1}{G_N} \int \mathrm{d} t \left( -a(t) \dot{a}(t)^2 -a(t) + \Lambda a(t)^3\right),
\label{miniact}
\end{equation}
where $G_N$ and $\Lambda$ denote Newton's constant and the cosmological constant in the continuum, {\it up to an overall sign}.
This overall sign is irrelevant when considering classical solutions $a(t)$, but it makes a crucial difference in a Euclidean path
integral whose weight factors are given by exp$(-S^{mini}_E)$. The action (\ref{miniact}) is unbounded below, due to the ``wrong'' sign
of the kinetic term, which implies that the path integral diverges for strongly oscillating scale factors $a(t)$. This is
a version of the infamous conformal-factor problem in gravity \cite{mamo,cure}, which is a consequence of the fact that the local conformal mode
of the metric appears in the Einstein-Hilbert action with a kinetic term of the wrong sign, rendering the action unbounded below,
irrespective of the signature of the four-metric.

In the phase space region where a stable, de Sitter-like vacuum state exists, the CDT path integral does not exhibit any conformal-factor
problem, so what has happened to the unboundedness of the Euclidean Einstein-Hilbert action in this formulation?
The answer is that the usual conformal mode contribution {\it is} present in the bare action, and does potentially lead to a divergence in the continuum limit.
However, for sufficiently small values of the bare inverse Newton coupling $\kappa_0$ this divergence is entropically suppressed
(see e.g. \cite{physrep} for a more detailed discussion). In other words, the collective contribution to the non-perturbative path integral
of the other, local degrees of freedom ``neutralizes'' this divergence. We can be more specific about what happens by examining
the minisuperspace action derived in CDT quantum gravity: it is of the same functional form as (\ref{miniact}), but
due to the entropic contributions it has the opposite overall sign.\footnote{Note that we are comparing the actions at fixed four-volume, where
$\Lambda$ is a Lagrange multiplier, in order to match the situation present in the Monte Carlo simulations.}
In this way, one obtains a de Sitter-like non-perturbative ground state for the full quantum theory, without tinkering with the Einstein-Hilbert
action or invoking a non-standard ``Wick rotation''.

We have come full circle from our considerations in Sec.\ \ref{intro}, where we emphasized the robustness of FLRW space-times
as effective models for classical cosmology. We then explained that there are obstacles to finding an analogous result
in the non-perturbative quantum theory, which take the form of phases of highly degenerate ``geometry'' without any apparent
classical limit.
CDT quantum gravity provides a way of keeping these instabilities in check
by imposing restrictions on the causal structure of the path integral configurations. Whether there
are alternative prescriptions that have the same effect is currently not known. However, once this problem is taken care of, the emergence of
a FLRW universe -- in this case a de Sitter universe -- appears to be more or less automatic.
How this particular quantum incarnation of the de Sitter universe is related to its classical counterparts in ``pure'' or ``real'' cosmology
(c.f. Fig.\ \ref{fig:GRQGcosmo}) needs to be analyzed in greater detail and clearly requires going beyond the global scale factor.
In the context of CDT, efforts are under way to define and implement quantum observables that can resolve more local
geometric properties, including curvature. Just like in the classical theory, it will be necessary to understand quantitatively how
a specific micro-structure contributes to observables on larger, coarse-grained scales.
Important information about how physics depends on the scale considered is usually captured by the renormalization group, but
how to implement it in non-perturbative quantum gravity is not immediately obvious. It is therefore encouraging that
there is a first, concrete proof of principle that renormalization group methods can be applied in CDT quantum gravity, despite its
background-free character \cite{renorm}. And perhaps there is more to be learned from comparing the ``averaging problems'' in
both the classical and quantum theories!

\begin{acknowledgements}
LG received funding from the People Programme (Marie Curie Actions) of the European Union's Seventh Framework ProgrammeFP7/2007-2013/ under REA grant agreement n. 706349 “Renormalisation Group methods for discrete Quantum Gravity”.
\end{acknowledgements}

%\section*{References}

\bibliography{ref}

\begin{thebibliography}{10}
\expandafter\ifx\csname url\endcsname\relax
  \def\url#1{\texttt{#1}}\fi
\expandafter\ifx\csname urlprefix\endcsname\relax\def\urlprefix{URL }\fi
\expandafter\ifx\csname href\endcsname\relax
  \def\href#1#2{#2} \def\path#1{#1}\fi

\bibitem{dark}
P.~Bull, et~al., {Beyond $\Lambda$CDM: Problems, solutions, and the road
  ahead}, Phys. Dark Univ. 12 (2016) 56--99.
\newblock \href {http://arxiv.org/abs/1512.05356} {\path{arXiv:1512.05356}},
  \href {http://dx.doi.org/10.1016/j.dark.2016.02.001}
  {\path{doi:10.1016/j.dark.2016.02.001}}.

\bibitem{challenges}
T.~Buchert, A.~A. Coley, H.~Kleinert, B.~F. Roukema, D.~L. Wiltshire,
  {Observational challenges for the standard FLRW model}, Int. J. Mod. Phys.
  D25~(03) (2016) 1630007.
\newblock \href {http://arxiv.org/abs/1512.03313} {\path{arXiv:1512.03313}},
  \href {http://dx.doi.org/10.1142/S021827181630007X}
  {\path{doi:10.1142/S021827181630007X}}.

\bibitem{inhomogeneous}
K.~Bolejko, M.~Korzynski, {Inhomogeneous cosmology and backreaction: current
  status and future prospects. }\href {http://arxiv.org/abs/1612.08222}
  {\path{arXiv:1612.08222}}.

\bibitem{as}
A.~Bonanno, F.~Saueressig, {Asymptotically safe cosmology - a status report.
  }\href {http://arxiv.org/abs/1702.04137} {\path{arXiv:1702.04137}}.

\bibitem{physrep}
J.~Ambj{\o}rn, A.~G{\"o}rlich, J.~Jurkiewicz, R.~Loll, {Nonperturbative quantum
  gravity}, Phys. Rept. 519 (2012) 127--210.
\newblock \href {http://arxiv.org/abs/1203.3591} {\path{arXiv:1203.3591}},
  \href {http://dx.doi.org/10.1016/j.physrep.2012.03.007}
  {\path{doi:10.1016/j.physrep.2012.03.007}}.

\bibitem{4dsimpl}
J.~Ambj{\o}rn, J.~Jurkiewicz, {Four-dimensional simplicial quantum gravity},
  Phys. Lett. B278 (1992) 42--50.
\newblock \href {http://dx.doi.org/10.1016/0370-2693(92)90709-D}
  {\path{doi:10.1016/0370-2693(92)90709-D}}.

\bibitem{point}
I.~Khavkine, R.~Loll, P.~Reska, {Coupling a point-like mass to quantum gravity
  with {Causal} {Dynamical} {Triangulations}}, Class. Quant. Grav. 27 (2010)
  185025.
\newblock \href {http://arxiv.org/abs/1002.4618} {\path{arXiv:1002.4618}},
  \href {http://dx.doi.org/10.1088/0264-9381/27/18/185025}
  {\path{doi:10.1088/0264-9381/27/18/185025}}.

\bibitem{wilson}
J.~Ambj{\o}rn, A.~G{\"o}rlich, J.~Jurkiewicz, R.~Loll, {Wilson loops in
  nonperturbative quantum gravity}, Phys. Rev. D92~(2) (2015) 024013.
\newblock \href {http://arxiv.org/abs/1504.01065} {\path{arXiv:1504.01065}},
  \href {http://dx.doi.org/10.1103/PhysRevD.92.024013}
  {\path{doi:10.1103/PhysRevD.92.024013}}.

\bibitem{regge}
T.~Regge, {General relativity without coordinates}, Nuovo Cim. 19 (1961)
  558--571.
\newblock \href {http://dx.doi.org/10.1007/BF02733251}
  {\path{doi:10.1007/BF02733251}}.

\bibitem{torus}
J.~Ambj{\o}rn, Z.~Drogosz, J.~Gizbert-Studnicki, A.~G{\"o}rlich, J.~Jurkiewicz,
  D.~Nemeth, {Impact of topology in causal dynamical triangulations quantum
  gravity}, Phys. Rev. D94~(4) (2016) 044010.
\newblock \href {http://arxiv.org/abs/1604.08786} {\path{arXiv:1604.08786}},
  \href {http://dx.doi.org/10.1103/PhysRevD.94.044010}
  {\path{doi:10.1103/PhysRevD.94.044010}}.

\bibitem{LCDT1}
S.~Jordan, R.~Loll, {Causal {Dynamical} {Triangulations} without preferred
  foliation}, Phys. Lett. B724 (2013) 155--159.
\newblock \href {http://arxiv.org/abs/1305.4582} {\path{arXiv:1305.4582}},
  \href {http://dx.doi.org/10.1016/j.physletb.2013.06.007}
  {\path{doi:10.1016/j.physletb.2013.06.007}}.

\bibitem{LCDT2}
S.~Jordan, R.~Loll, {De {Sitter} universe from Causal Dynamical Triangulations
  without preferred foliation}, Phys. Rev. D88 (2013) 044055.
\newblock \href {http://arxiv.org/abs/1307.5469} {\path{arXiv:1307.5469}},
  \href {http://dx.doi.org/10.1103/PhysRevD.88.044055}
  {\path{doi:10.1103/PhysRevD.88.044055}}.

\bibitem{cdt2001}
J.~Ambj{\o}rn, J.~Jurkiewicz, R.~Loll, {Dynamically triangulating Lorentzian
  quantum gravity}, Nucl. Phys. B610 (2001) 347--382.
\newblock \href {http://arxiv.org/abs/hep-th/0105267}
  {\path{arXiv:hep-th/0105267}}, \href
  {http://dx.doi.org/10.1016/S0550-3213(01)00297-8}
  {\path{doi:10.1016/S0550-3213(01)00297-8}}.

\bibitem{reconstructing}
J.~Ambj{\o}rn, J.~Jurkiewicz, R.~Loll, {Reconstructing the universe}, Phys.
  Rev. D72 (2005) 064014.
\newblock \href {http://arxiv.org/abs/hep-th/0505154}
  {\path{arXiv:hep-th/0505154}}, \href
  {http://dx.doi.org/10.1103/PhysRevD.72.064014}
  {\path{doi:10.1103/PhysRevD.72.064014}}.

\bibitem{horava}
P.~Ho{\v r}ava, {Quantum gravity at a Lifshitz point}, Phys. Rev. D79 (2009)
  084008.
\newblock \href {http://arxiv.org/abs/0901.3775} {\path{arXiv:0901.3775}},
  \href {http://dx.doi.org/10.1103/PhysRevD.79.084008}
  {\path{doi:10.1103/PhysRevD.79.084008}}.

\bibitem{cdthl}
J.~Ambj{\o}rn, A.~G{\"o}rlich, S.~Jordan, J.~Jurkiewicz, R.~Loll, {CDT} meets
  {Horava}-{Lifshitz} gravity, Phys. Lett. B690 (2010) 413--419.
\newblock \href {http://dx.doi.org/10.1016/j.physletb.2010.05.054}
  {\path{doi:10.1016/j.physletb.2010.05.054}}.

\bibitem{characteristics}
J.~Ambj{\o}rn, J.~Gizbert-Studnicki, A.~G{\"o}rlich, J.~Jurkiewicz,
  N.~Klitgaard, R.~Loll, {Characteristics of the new phase in CDT}, Eur. Phys.
  J. C77~(3) (2017) 152.
\newblock \href {http://arxiv.org/abs/1610.05245} {\path{arXiv:1610.05245}},
  \href {http://dx.doi.org/10.1140/epjc/s10052-017-4710-3}
  {\path{doi:10.1140/epjc/s10052-017-4710-3}}.

\bibitem{emergence}
J.~Ambj{\o}rn, J.~Jurkiewicz, R.~Loll, {Emergence of a 4-{D} world from causal
  quantum gravity}, Phys. Rev. Lett. 93 (2004) 131301.
\newblock \href {http://arxiv.org/abs/hep-th/0404156}
  {\path{arXiv:hep-th/0404156}}, \href
  {http://dx.doi.org/10.1103/PhysRevLett.93.131301}
  {\path{doi:10.1103/PhysRevLett.93.131301}}.

\bibitem{transition1}
J.~Ambj{\o}rn, S.~Jordan, J.~Jurkiewicz, R.~Loll, {A second-order phase
  transition in CDT}, Phys. Rev. Lett. 107 (2011) 211303.
\newblock \href {http://arxiv.org/abs/1108.3932} {\path{arXiv:1108.3932}},
  \href {http://dx.doi.org/10.1103/PhysRevLett.107.211303}
  {\path{doi:10.1103/PhysRevLett.107.211303}}.

\bibitem{transition2}
J.~Ambj{\o}rn, S.~Jordan, J.~Jurkiewicz, R.~Loll, {Second- and {first}-{order}
  {phase} {transitions} in {CDT}}, Phys. Rev. D85 (2012) 124044.
\newblock \href {http://arxiv.org/abs/1205.1229} {\path{arXiv:1205.1229}},
  \href {http://dx.doi.org/10.1103/PhysRevD.85.124044}
  {\path{doi:10.1103/PhysRevD.85.124044}}.

\bibitem{effective}
J.~Ambj{\o}rn, J.~Gizbert-Studnicki, A.~G{\"o}rlich, J.~Jurkiewicz, {The
  effective action in 4-dim {CDT}. {The} transfer matrix approach}, JHEP 06
  (2014) 034.
\newblock \href {http://arxiv.org/abs/1403.5940} {\path{arXiv:1403.5940}},
  \href {http://dx.doi.org/10.1007/JHEP06(2014)034}
  {\path{doi:10.1007/JHEP06(2014)034}}.

\bibitem{signature}
J.~Ambj{\o}rn, D.~N. Coumbe, J.~Gizbert-Studnicki, J.~Jurkiewicz, {Signature
  change of the metric in {CDT} quantum gravity?}, JHEP 08 (2015) 033.
\newblock \href {http://arxiv.org/abs/1503.08580} {\path{arXiv:1503.08580}},
  \href {http://dx.doi.org/10.1007/JHEP08(2015)033}
  {\path{doi:10.1007/JHEP08(2015)033}}.

\bibitem{newphasetransition}
D.~N. Coumbe, J.~Gizbert-Studnicki, J.~Jurkiewicz, {Exploring the new phase
  transition of {CDT}}, JHEP 02 (2016) 144.
\newblock \href {http://arxiv.org/abs/1510.08672} {\path{arXiv:1510.08672}},
  \href {http://dx.doi.org/10.1007/JHEP02(2016)144}
  {\path{doi:10.1007/JHEP02(2016)144}}.

\bibitem{semilimit}
J.~Ambj{\o}rn, A.~G{\"o}rlich, J.~Jurkiewicz, R.~Loll, J.~Gizbert-Studnicki,
  T.~Trzesniewski, {The semiclassical limit of {Causal} {Dynamical}
  {Triangulations}}, Nucl. Phys. B849 (2011) 144--165.
\newblock \href {http://arxiv.org/abs/1102.3929} {\path{arXiv:1102.3929}},
  \href {http://dx.doi.org/10.1016/j.nuclphysb.2011.03.019}
  {\path{doi:10.1016/j.nuclphysb.2011.03.019}}.

\bibitem{transfer}
J.~Ambj{\o}rn, J.~Gizbert-Studnicki, A.~G{\"o}rlich, J.~Jurkiewicz, The
  {transfer} matrix in four-dimensional {CDT}, JHEP 1209 (2012) 017.
\newblock \href {http://dx.doi.org/10.1007/JHEP09(2012)017}
  {\path{doi:10.1007/JHEP09(2012)017}}.

\bibitem{measure}
J.~Ambj{\o}rn, L.~Glaser, A.~G{\"o}rlich, J.~Jurkiewicz, {Euclidian 4d quantum
  gravity with a non-trivial measure term}, JHEP 10 (2013) 100.
\newblock \href {http://arxiv.org/abs/1307.2270} {\path{arXiv:1307.2270}},
  \href {http://dx.doi.org/10.1007/JHEP10(2013)100}
  {\path{doi:10.1007/JHEP10(2013)100}}.

\bibitem{4dEDT}
J.~Ambj{\o}rn, J.~Jurkiewicz, {Scaling in four-dimensional quantum gravity},
  Nucl. Phys. B451 (1995) 643--676.
\newblock \href {http://arxiv.org/abs/hep-th/9503006}
  {\path{arXiv:hep-th/9503006}}, \href
  {http://dx.doi.org/10.1016/0550-3213(95)00303-A}
  {\path{doi:10.1016/0550-3213(95)00303-A}}.

\bibitem{3dEDT}
J.~Ambjorn, S.~Varsted, {Three-dimensional simplicial quantum gravity}, Nucl.
  Phys. B373 (1992) 557--577.
\newblock \href {http://dx.doi.org/10.1016/0550-3213(92)90444-G}
  {\path{doi:10.1016/0550-3213(92)90444-G}}.

\bibitem{5dEDT}
A.~George, {Five-dimensional dynamical triangulations}, Ph.D. thesis, Swansea
  U. (1999).
\newblock \href {http://arxiv.org/abs/hep-lat/9909033}
  {\path{arXiv:hep-lat/9909033}}.

\bibitem{10dEDT}
A.~I. Veselov, M.~A. Zubkov, {10-D Euclidean quantum gravity}, Phys. Lett. B591
  (2004) 311.
\newblock \href {http://arxiv.org/abs/hep-lat/0306030}
  {\path{arXiv:hep-lat/0306030}}, \href
  {http://dx.doi.org/10.1016/j.physletb.2004.04.047}
  {\path{doi:10.1016/j.physletb.2004.04.047}}.

\bibitem{melon}
R.~Gurau, J.~P. Ryan, {Melons are branched polymers}, Annales Henri Poincare
  15~(11) (2014) 2085--2131.
\newblock \href {http://arxiv.org/abs/1302.4386} {\path{arXiv:1302.4386}},
  \href {http://dx.doi.org/10.1007/s00023-013-0291-3}
  {\path{doi:10.1007/s00023-013-0291-3}}.

\bibitem{firstcdt}
J.~Ambj{\o}rn, R.~Loll, {Nonperturbative {Lorentzian} quantum gravity,
  causality and topology change}, Nucl. Phys. B536 (1998) 407--434.
\newblock \href {http://arxiv.org/abs/hep-th/9805108}
  {\path{arXiv:hep-th/9805108}}, \href
  {http://dx.doi.org/10.1016/S0550-3213(98)00692-0}
  {\path{doi:10.1016/S0550-3213(98)00692-0}}.

\bibitem{desitter1}
J.~Ambj{\o}rn, A.~G{\"o}rlich, J.~Jurkiewicz, R.~Loll, {Planckian birth of the
  quantum de {Sitter} universe}, Phys. Rev. Lett. 100 (2008) 091304.
\newblock \href {http://arxiv.org/abs/0712.2485} {\path{arXiv:0712.2485}},
  \href {http://dx.doi.org/10.1103/PhysRevLett.100.091304}
  {\path{doi:10.1103/PhysRevLett.100.091304}}.

\bibitem{desitter2}
J.~Ambj{\o}rn, A.~G{\"o}rlich, J.~Jurkiewicz, R.~Loll, {The nonperturbative
  quantum de {Sitter} universe}, Phys. Rev. D78 (2008) 063544.
\newblock \href {http://arxiv.org/abs/0807.4481} {\path{arXiv:0807.4481}},
  \href {http://dx.doi.org/10.1103/PhysRevD.78.063544}
  {\path{doi:10.1103/PhysRevD.78.063544}}.

\bibitem{spectral}
J.~Ambj{\o}rn, J.~Jurkiewicz, R.~Loll, {Spectral dimension of the universe},
  Phys. Rev. Lett. 95 (2005) 171301.
\newblock \href {http://arxiv.org/abs/hep-th/0505113}
  {\path{arXiv:hep-th/0505113}}, \href
  {http://dx.doi.org/10.1103/PhysRevLett.95.171301}
  {\path{doi:10.1103/PhysRevLett.95.171301}}.

\bibitem{carlip}
S.~Carlip, {Spontaneous dimensional reduction in quantum gravity}, Int. J. Mod.
  Phys. D25~(12) (2016) 1643003.
\newblock \href {http://arxiv.org/abs/1605.05694} {\path{arXiv:1605.05694}},
  \href {http://dx.doi.org/10.1142/S0218271816430033}
  {\path{doi:10.1142/S0218271816430033}}.

\bibitem{thermal}
G.~Amelino-Camelia, F.~Brighenti, G.~Gubitosi, G.~Santos, {Thermal dimension of
  quantum spacetime}, Phys. Lett. B767 (2017) 48--52.
\newblock \href {http://arxiv.org/abs/1602.08020} {\path{arXiv:1602.08020}},
  \href {http://dx.doi.org/10.1016/j.physletb.2017.01.050}
  {\path{doi:10.1016/j.physletb.2017.01.050}}.

\bibitem{searching}
J.~Ambj{\o}rn, D.~Coumbe, J.~Gizbert-Studnicki, J.~Jurkiewicz, {Searching for a
  continuum limit in {Causal} {Dynamical} {Triangulation} quantum gravity},
  Phys. Rev. D93~(10) (2016) 104032.
\newblock \href {http://arxiv.org/abs/1603.02076} {\path{arXiv:1603.02076}},
  \href {http://dx.doi.org/10.1103/PhysRevD.93.104032}
  {\path{doi:10.1103/PhysRevD.93.104032}}.

\bibitem{semiclassical}
J.~Ambj{\o}rn, J.~Jurkiewicz, R.~Loll, {Semiclassical universe from first
  principles}, Phys. Lett. B607 (2005) 205--213.
\newblock \href {http://arxiv.org/abs/hep-th/0411152}
  {\path{arXiv:hep-th/0411152}}, \href
  {http://dx.doi.org/10.1016/j.physletb.2004.12.067}
  {\path{doi:10.1016/j.physletb.2004.12.067}}.

\bibitem{geometry}
J.~Ambj{\o}rn, A.~G{\"o}rlich, J.~Jurkiewicz, R.~Loll, {Geometry of the quantum
  universe}, Phys. Lett. B690 (2010) 420--426.
\newblock \href {http://arxiv.org/abs/1001.4581} {\path{arXiv:1001.4581}},
  \href {http://dx.doi.org/10.1016/j.physletb.2010.05.062}
  {\path{doi:10.1016/j.physletb.2010.05.062}}.

\bibitem{entropy}
J.~Ambj{\o}rn, A.~G{\"o}rlich, J.~Jurkiewicz, R.~Loll, {CDT - an entropic
  theory of quantum gravity}, in: {Proceedings, Workshop on Continuum and
  lattice approaches to quantum gravity (CLAQG08): Brighton, UK, Sep 17-19,
  2008}.
\newblock \href {http://arxiv.org/abs/1007.2560} {\path{arXiv:1007.2560}}.

\bibitem{haha}
J.~B. Hartle, S.~W. Hawking, {Wave function of the universe}, Phys. Rev. D28
  (1983) 2960--2975.
\newblock \href {http://dx.doi.org/10.1103/PhysRevD.28.2960}
  {\path{doi:10.1103/PhysRevD.28.2960}}.

\bibitem{mamo}
P.~O. Mazur, E.~Mottola, {The gravitational measure, solution of the conformal
  factor problem and stability of the ground state of quantum gravity}, Nucl.
  Phys. B341 (1990) 187--212.
\newblock \href {http://dx.doi.org/10.1016/0550-3213(90)90268-I}
  {\path{doi:10.1016/0550-3213(90)90268-I}}.

\bibitem{cure}
A.~Dasgupta, R.~Loll, {A proper time cure for the conformal sickness in quantum
  gravity}, Nucl. Phys. B606 (2001) 357--379.
\newblock \href {http://arxiv.org/abs/hep-th/0103186}
  {\path{arXiv:hep-th/0103186}}, \href
  {http://dx.doi.org/10.1016/S0550-3213(01)00227-9}
  {\path{doi:10.1016/S0550-3213(01)00227-9}}.

\bibitem{renorm}
J.~Ambj{\o}rn, A.~G{\"o}rlich, J.~Jurkiewicz, A.~Kreienbuehl, R.~Loll,
  {Renormalization group flow in {CDT}}, Class. Quant. Grav. 31 (2014) 165003.
\newblock \href {http://arxiv.org/abs/1405.4585} {\path{arXiv:1405.4585}},
  \href {http://dx.doi.org/10.1088/0264-9381/31/16/165003}
  {\path{doi:10.1088/0264-9381/31/16/165003}}.

\end{thebibliography}

\end{document}